\documentclass[10pt,twocolumn,twoside]{IEEEtran}  % To submit to IEEE (10pt, double column, single spaced)
\normalsize

\usepackage{ifpdf}
\usepackage{cite}
\usepackage{graphicx,color}
\usepackage{epstopdf}
\usepackage{amsmath, amsfonts, amsmath, amssymb,amsxtra, mathrsfs, pifont}

\definecolor{dgreen}{rgb}{0,0.5,0}
\begin{document}
%\pagenumbering{gobble}
%
\title{Unitary Checkerboard Precoded OFDM for Low-PAPR Optical Wireless Communications}

\markboth{}{Submitted paper}

\author{
	\IEEEauthorblockN{  Traian~E.~Abrudan,\IEEEauthorrefmark{1}, %\\
                        %Dong %Fang,~\IEEEmembership{Member,~IE%EE,}\IEEEauthorrefmark{2} % - Dong's message in Linked in was to exclude his name from the paper
                        Stepan Kucera,\IEEEauthorrefmark{2} %\\
                        Holger Claussen\IEEEauthorrefmark{3}
                        } \\
    \IEEEauthorblockA{\IEEEauthorrefmark{1}Nokia Bell Labs, Espoo, Finland.}\\
    \IEEEauthorblockA{\IEEEauthorrefmark{2}Nokia Bell Labs, Munich, Germany.} \\
    \IEEEauthorblockA{\IEEEauthorrefmark{3}Tyndall National Institute, Dublin, Ireland.
    }
		%\thanks{Manuscript
		%	received xxxxxx XX, 2018;
		%	revised xxxxxx XX, 2018,
		%	accepted xxxxxx XX, 2018.
		%	%The authors would like to thank...
                %}
}

\maketitle

%%%%%%%%%%%%%%%%%%%%%%%%%%%%%%%%%%%%%%%%%%%%%%%%%%%%%%%%%%%%%%%%%%%%%%%%%%%%%%%%%%%%%%%%%%%%%%%%%%%%%%%%%%%%%%%%%%%%%%%%%%%%%%%%%%%%%%%%%%%%%%%%%
	
\begin{abstract}
  
  %\revcom{Make connection for to 6G and cite this paper:"} \cite{2021_WanHua_etal}.	
  Future 6G wireless networks will once again have to raise the capability in most of the technology domains by a factor of 10-100. Depending on the application, future requirements include peak data rates of 1Tb/s per user, 0.1ms latency, less than 1 out of a million outage, centimetre accurate positioning, near zero energy consumption at the device, and operation in different environments including factories, vehicles, and more \cite{2021_UusRugBol_etal, 2020_Viswanathan, 2021_WanHua_etal}. Optical wireless communications (OWC) have the potential to provide ultra-high data rates in a cost effective way, thanks to the vast and freely available light spectrum, and the availability of devices for transmitters and receivers. 5G NR architecture permits the integration of stand-alone OWC nodes on network layer~\cite{3GPP_Rel15}. Current 6G research investigates advanced physical layer designs including OWC-compatible waveforms. In this context, in this paper a new pre-coded orthogonal frequency division multiplexing (OFDM) waveform is proposed that is tailored to the OWC specific needs.
  Its prime advantage compared to OFDM is the ultra-low peak-to-average power ratio (PAPR), while preserving other benefits,
  such as high spectral efficiency, flexible subcarrier nulling, and low computational complexity.

\end{abstract}

	
\maketitle

%%%%%%%%%%%%%%%%%%%%%%%%%%%%%%%%%%%%%%%%%%%%%%%%%%%%%%%%%%%%%%%%%%%%%%%%%%%%%%%%%%%%%%%%%%%%%%%%%%%%%%%%%%%%%%%%%%%%%%%%%%%%%%%%%%%%%%%%%%%%%%%%%

% \fi
\IEEEpeerreviewmaketitle

%{Intro}
\section{Introduction \label{sec:intro}}
%%%%%%%%%%%%%%%%%%%%%%%%%%%%%%%%%%%%%%%%%%%%%%%%%%%%%%%%%%%%%%%%%%%%%%%%%%%%%%%%%%%%%%%%%%%%%%%%%%%%%%%%%%%%%%%%%%%%%%%%%%%%%%%%%%%%%%%%%%%%%%%%%

%\IEEEPARstart{L}{ighting} 
The lighting industry currently is undergoing a transformation from simple bulbs and fluorescent tubes to intelligent solid state LED lighting systems (luminaires).
To enable remote lighting control, the luminaires are connected and even powered over Ethernet.
An ultra-dense radio access deployment can be thus realized at very low cost through embedding wireless access units into every luminaire, supporting for example Wi-Fi radio access.
The synergy between the existing luminaire infrastructure and wireless communications can be maximized by reusing also the visible light spectrum in which luminaires operate for wireless data transfers~\cite{2015_HuaWanShiWanChi, 2013_SteSiuZwi, 2012_CosKhaChoCorCia}.
Herein, the drivers of the LED light sources are directly modulated with payload data to deliver wireless data to nearby user devices equipped with an optical receiver.

Optical light communications (OWC) have been demonstrated to achieve wireless data rates in the order of Gigabits per second
\cite{2017_IslFerHe_etal, 2014_CosWajCorCia, 2015_TsoVidHaa, 2015_HusElm, 2015_RamRetMak_etal, 2014_TsoChuRajKen_etal, 2012_CosKhaChoCorCia}
by modulating the intensity of a light source~\cite{2015_TsoVidHaa}.
On the physical layer, orthogonal frequency division multiplexing (OFDM) is seen as the most promising techniques due to its high spectral efficiency,
excellent multipath resilience and simple channel estimation/equalization, already proven in 4G and 5G radio frequency systems.

Before we introduce the proposed modulation scheme, we review {\em three key characteristics of OWC.}
%a
\begin{itemize}
\item[1)] {\em Baseband signal constraints:}
OWC poses two specific constraints on the baseband signal, which are not characteristic to RF-based communications~\cite{2016_IslHaa,2014_TsoVidHaa}.
Unlike radio frequency (RF) carriers, light intensity modulation bears no phase information, and therefore, the transmitted OWC signal must be {\em real-valued} and {\em positive}. These two contraints have the folowing consequences:
i) it halves the spectral efficiency compared to complex-valued signals used in RF. Specifically, in the case of OFDM modulation, the real-valued constraint on the signal implies Hermitian-symmetry of the symbols ~\cite{2014_TsoVidHaa}.
ii) since an {\em even} FFT size is used in practice (for computational efficiency reasons),
the subcarrier at the lower edge of the spectrum must be null.
iii) the limitation of the OWC signal to positive values only halves the signal dynamic range compared to RF. For this reason, Direct Current (DC) biasing~\cite{1996_CarKah,2014_TsoVidHaa} or unipolar modulations~\cite{2006_ArmLow,2012_TsoSinHaa,2015_TsoVidHaa} are typically used.
\item[2)] {\em Front-end limitations:}
High PAPR of the OFDM signal poses serious constraints on the OWC optical front-end design.
Light sources, typically light emission diodes (LEDs) or laser diodes, exhibit high non-linearities which limit the signal dynamic range to a small quasi-linear region~\cite{2015_LiMajHenHaa,2020_CuiZubCol}.
Ambient light dimming (adjusting the illumination level) further reduces the dynamic range of the OFDM signal, thus high PAPR may result in signal clipping~\cite{2016_IslHaa,2014_TsoVidHaa}.
\item[3)] {\em Low-frequency disturbances:}
Due to time-varying ambient light conditions (especially in mobile scenarios), interference from fluorescent light, as well as AC-coupling in the transceiver,
low frequencies should be avoided~\cite{2013_AzhBri,2016_GroJunLanHaaWol,2016_LinHuaJi,2014_TsoVidHaa}.
Hence, the lower OFDM subcarrier(s) do not carry data, i.e., they are {\em null subcarriers}.
\end{itemize}
This paper proposes an innovative OFDM-based modulation scheme characterized by four distinct OWC-enabling features:
\begin{itemize}
\item  {\em Ultra-low PAPR:}
The proposed scheme achieves ultra-low PAPR levels, comparable to baseband modulation.
\item {\em High spectral efficiency:}
  Under the three key OWC constraints listed previously, and other design constraints -- such as the use of cyclic prefix that tremendously simplifies channel estimation and equalization --
  the proposed modulation scheme fully utilizes the available bandwidth.
\item {\em Flexible subcarrier nulling:}
Our scheme is capable of arbitrarily nulling subcarriers i.e., in order to
eliminate low frequency disturbances. %, and/or to enable multi-user access. 
Note that the presence of null subcarriers is {\em not} imposed by our scheme, it is rather a practical design requirement.
\item {\em  Low implementation complexity}:
Low computational complexity of modulation/demodulation is a key precursor for cost-efficient delivery of high data rates.
The proposed scheme exhibits similar complexity with the conventional OFDM transceiver.
It only requires simple operations such as FFT/IFFT and multiplication by low-rank matrices both at TX and RX, thus making it appealing for very high-speed OWC.
\end{itemize}

More specifically, a precoded OFDM scheme referred to as the {\em Unitary Checkerboard Precoded OFDM (UCP-OFDM)} is proposed whose main idea is to use specially designed linear precoding before the Inverse Fast Fourier Transform (IFFT) operation at the transmitter to achieve ultra-low PAPR of the transmitted OFDM signal. Simulations demonstrate PAPR levels just 1-2 dB above the baseband modulation scheme which may be considered to be a practical PAPR lower bound.

However, unlike the baseband (BB) modulation, the proposed UCP-OFDM modulation also provides full control over the presence of null subcarriers, regardless of their location and purpose, such as elimination of low-frequency disturbances. %, or multi-user access across subcarriers. 
Since the proposed precoder is a unitary transform (hence full-rank matrix),
it also preserves the noise covariance (no noise coloring takes place at the receiver), and is distortionless (in absence of noise, the data symbols are recovered exactly).
In addition, the proposed scheme is non-redundant, hence it achieves full throughput. Furthermore, the proposed precoder is data-independent, unlike other schemes that either introduce overhead, thus decreasing the data rate, or require expensive optimization to be carried out in real-time for every transmitted OFDM symbol in order to either shuffle null subcarriers, %~\cite{2011_TsoSinHaa},
or calculate the redundant part of the symbol. 

In practical terms, the implementation complexity of the proposed scheme is comparable to the conventional OFDM which may be considered as the simplest multipath-resilient modulation known to date, that benefits from trivial single-tap channel estimation and equalization.

The paper is organized as follows. A summary of commonly used OWC schemes is given in Section \ref{sec:sota}. After defining the system model in Section \ref{sec:sys}, the proposed unitary checkerboard precoded OFDM scheme is explained in more detail in Section \ref{sec:precoder}.
Implementation issues discussed in Section \ref{implementation}. Numerical evaluations and conclusions follow in Section \ref{evaluation} and \ref{sec:conclusion}, respectively.

%%%%%%%%%%%%%%%%%%%%%%%%%%%%%%%%%%%%%%%%%%%%%%%%%%%%%%%%%%%%%%%%%%%%%%%%%%%%%%%%%%%%%%%%%%%%%%%%%%%%%%%%%%%%%%%%%%%%%%%%%%%%%%%%%%%%%%%%%%%%%%%%%
\section{Overview of Modulation Schemes For OWC \label{sec:sota}}
%%%%%%%%%%%%%%%%%%%%%%%%%%%%%%%%%%%%%%%%%%%%%%%%%%%%%%%%%%%%%%%%%%%%%%%%%%%%%%%%%%%%%%%%%%%%%%%%%%%%%%%%%%%%%%%%%%%%%%%%%%%%%%%%%%%%%%%%%%%%%%%%%

The PAPR reduction problem has been extensively studied in the literature in the context of RF communication. For example Tellado~\cite{2000_Tel}, summarized various PAPR reduction techniques such as: 
i) clipping methods, which is essentially OFDM with amplitude limitation, ii) tone reservation method, which assigns specific tones for transmitting control symbols, hence sacrificing spectral efficiency, iii) tone injection method which relies on cyclic constellation expansion, hence it requires heavy optimization to be carried out, iv) selected mapping schemes~\cite{2010_LiWanWan} which can be regarded as redundant coding, which again, decreases the spectral efficiency. For these reasons, with the constrains 1)-4) in Section~\ref{sec:intro} in mind, various modulation schemes have been proposed in the OWC literature (a comprehensive classification is given in~\cite{2016_IslHaa}), but none of them achieves full throughout and permits on-demand subcarrier nulling simultaneously, subject to low PAPR, and computationally efficient modulation/demodulation.
Challenges and potential of OWC have been reviewed in~\cite{2010_MinGhaBriFau, 2011_ElgMesHaa, 2015_PatFenHuMoh, 2015_LiMajHenHaa, 2020_CuiZubCol, 2020_MirGuzGalGiu}, and comprehensive literature surveys are provided
in~\cite{2015_KarZafKalPar, 2011_ElgMesHaa, 2015_PatFenHuMoh}. Below, we summarize the most relevant OWC modulations:

{\em DC-Offset OFDM (DCO-OFDM)} as proposed in \cite{1996_CarKah} is perhaps the most common OWC modulation.
DCO-OFDM optimizes the DC bias value in order to minimize the signal clipping distortions, rather than reducing the high PAPR level of the OFDM signal. Therefore, the PAPR levels remain high.

{\em Asymmetrically Clipped Optical OFDM (ACO-OFDM)} \cite{2006_ArmLow} uses a zero DC bias value to obtain a positive time-domain signal.
Negative samples are all set to zero, and only odd subcarriers are modulated.
This way, most clipping noise becomes orthogonal to the desired signal, i.e., it appears on even subcarriers.
However, due to halving the number of active subcarriers, the spectral efficiency is further reduced by half.

{\em Discrete Hartley Transform (DHT)} modulation has been proposed in \cite{2010_MorMunJun} to achieve lower PAPR.
When a real-valued constellation is used, the method produces a real-valued signal.
However, it also reduces spectral efficiency by half since it involves asymmetric clipping used in ACO-OFDM.

{\em Carrierless Amplitude Phase (CAP)} modulation method was proposed in~\cite{2012_WuLinWei_etal} to achieve lower PAPR than multicarrier modulations.
CAP modulation involves a pair of Hilbert filters that produce two time-domain orthogonal sequences corresponding to the in-phase and quadrature signals.
However, CAP does not benefit from the great advantage of OFDM in dealing with multipath (single-tap equalization),
and requires very complex nonlinear equalizers such as Volterra filters~\cite{2015_WanTaoHuaShiChi, 2015_SteMakSiu, 2013_SteSiuZwi}.

{\em Baseband (BB)} modulation with frequency-domain equalization (FDE) possesses very low PAPR.
This includes also the the FFT-precoded OFDM used in the 4G Long Term Evolution and 5G New Radio wireless communication standards. Pulse amplitude modulations (PAM) schemes produce real-valued signals, but they are severely affected by baseline wander (due to the loss of DC-component) which degrade the overall bit-rate performance, as shown in~\cite{2016_GroJunLanHaaWol}.
This limits its use to low-order modulations only, hence it achieves low spectral efficiency~\cite{2016_GroJunLanHaaWol}.
In order to circumvent this issue up-conversion may be employed as in~\cite{2014Wan_etal}, which requires additional oscillators and filters,
thus increasing the transceiver complexity.
Other BB schemes use line codes in conjunction with pre-emphasis techniques in order to cope with signal baseline wander and data-dependent jitter~\cite{2016_LinHuaJi},
but they do not achieve the high spectral efficiency of OFDM.

{\em Unipolar-OFDM (U-OFDM)} technique proposed in \cite{2012_TsoSinHaa} achieves a positive time-domain signal by repeating the bipolar OFDM signal samples twice,
multiplying the second half by -1, and then setting all negative samples to zero.
This way, no biasing is required, and clipping is asymmetrical.
Yet as a consequence of double symbol length, spectral efficiency is reduced by half, similarly to ACO-OFDM.
Additional spectral efficiency loss is caused by both positive and negative halves requiring separate cyclic prefixes.

{\em Enhanced Unipolar OFDM (eU-OFDM)}, proposed in \cite{2014_TsoHaa}, attempts to compensate for the spectral efficiency loss of U-OFDM by superimposing several unipolar streams.
However, decoding is very complex, since it is applied to very large data blocks whose length increases exponentially (powers of two) with the number of superimposed streams.

{\em Enhanced Asymmetrically Clipped Optical OFDM (eACO-OFDM)} \cite{2015_IslTsoHaa} has been recently introduced to compensate for the spectral efficiency loss of ACO-OFDM,
by superimposing ACO multiple data streams. Unlike in eU-OFDM, processing takes place in short blocks, equal to the OFDM symbols length, hence it its complexity is lower.
Both eU-OFDM and eACO-OFDM employ very complex successive interference cancellation (SIC), i.e., the decoded streams need to be re-encoded and pulse-shaped, and then
they undergo again channel distortion in order to perform interference subtraction.
Due to SIC decoding, eU-OFDM and eACO-OFDM suffer from error propagation between subsequent decoding stages.

{\em Coded Schemes}
More recently, Single Carrier Generalized Time Slot Index Modulation (SC-GTIM)~\cite{2020_PurYesSafHaa} and Layered Asymmetrically Clipped Optical OFDM (LACO-OFDM)~\cite{2021_ZhaZunPetHaa} that combine modulation and coding have been proposed. 
SC-GTIM encodes the information on the amplitudes and positions of pulses and then, a set partitioning algorithm is proposed to construct a codebook. LACO-OFDM bowwrows design features from both ACO-OFDM and DCO-OFDM combined with forward error correction coding.

%%%%%%%%%%%%%%%%%%%%%%%%%%%%%%%%%%%%%%%%%%%%%%%%%%%%%%%%%%%%%%%%%%%%%%%%%%%%%%%%%%%%%%%%%%%%%%%%%%%%%%%%%%%%%%%%%%%%%%%%%%%%%%%%%%%%%%%%%%%%%%%%%
\section{System Model \label{sec:sys}}
%%%%%%%%%%%%%%%%%%%%%%%%%%%%%%%%%%%%%%%%%%%%%%%%%%%%%%%%%%%%%%%%%%%%%%%%%%%%%%%%%%%%%%%%%%%%%%%%%%%%%%%%%%%%%%%%%%%%%%%%%%%%%%%%%%%%%%%%%%%%%%%%%

We assume a multicarrier transmission with $N$ subcarriers of which $M$ are active (hence there are $Z=N-M$ null subcarriers).
Next, transmitter (TX) and receiver (RX) architectures and functionalities are explained in detail.

\begin{figure*}[t]
    \centering
    \includegraphics[width=0.7\textwidth]{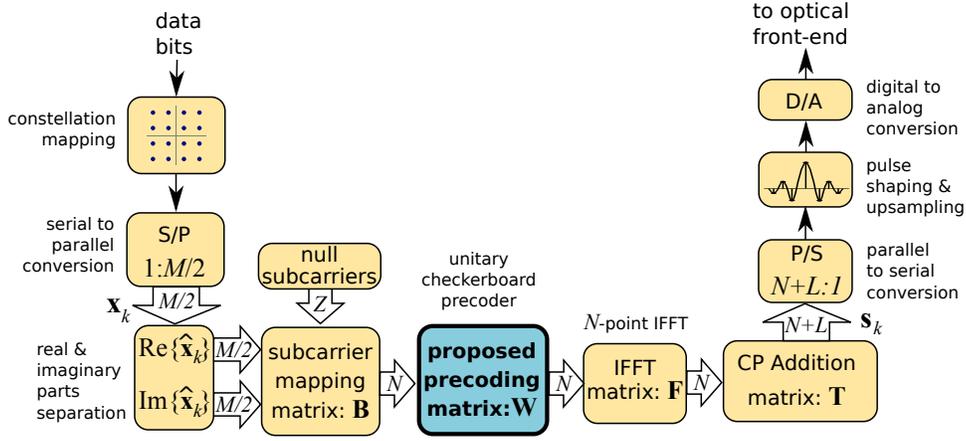}
    \caption{TX block diagram. Proposed linear precoding (highlighted block) takes place before the IFFT operation. DC biasing takes place at the optical front-end.}
    \label{fig:TX}
\end{figure*}

\subsection{Transmitter Architecture}

The transmitter (TX) block diagram is shown in Figure~\ref{fig:TX}.
A coded data bit stream is sequentially mapped to blocks of complex-valued constellation points\footnote{we chose the adopt this ``complexification'' approach in order to be able to straight-forwardly use
some well-established transceiver function such as Gray mapping of QAM constellation symbols to bits, trellis-coded modulation, etc.}
represented by $M/2 \times 1$ column vectors ${\mathbf x}_k.$
The $M/2$ real parts and the $M/2$ imaginary parts of ${\mathbf x}_k$ are then used to construct $M \times 1$ real-valued vectors,
which essentially consist of PAM (Pulse Amplitude Modulation) levels to be sent across $M$ active subcarriers.
%Existing multicarrier OWC schemes typically impose Hermitian symmetry of the frequency domain symbols in order to produce a real-valued signal, whereas
The proposed scheme outputs a real-valued signal as long as its input is a real-valued signal (i.e., the real and imaginary parts of ${\mathbf x}_k$).
$Z$ zeros corresponding to the null subcarriers are then inserted at predefined subcarrier indices to form  $N \times 1$ vectors,
operation performed by the subcarrier mapping matrix ${\mathbf B}.$

The resulting signal vector is input to the proposed precoder ${\mathbf W}$ (the highlighted block in Figure~\ref{fig:TX}).
The precoder will be described in detail in Section~\ref{sec:precoder}.
After precoding, the usual IFFT operation is performed, followed by adding a cyclic prefix (CP) of length $L$.
The signal is then serialized, oversampled and pulse-shaped, and then converted to analog waveform for the optical front-end.
Then, DC biasing, pre-equalization and amplification are performed before the signal is converted to optical signal.

The overall signal processing along the horizontal part of the TX chain in Figure~\ref{fig:TX} may be described by a simple matrix equation
which expresses the time-domain precoded OFDM samples as follows
\begin{equation}
{\mathbf s}_k={\mathbf T}{\mathbf F}{\mathbf W}{\mathbf B}
\left[\!\!
\begin{array}{cc}
{\mathrm{Re}}\{{\mathbf x_k}\} \\
{\mathrm{Im}}\{{\mathbf x_k}\} \\
\end{array}
\!\!\right],
\end{equation}
where ${\mathbf T}$ is the $(N+L) \times N$ CP addition matrix, which is formed by last $L$ rows of ${\mathbf I}_N$ followed by all the rows of ${\mathbf I}_N$ itself. ${\mathbf F}$ is the $N \times N$ power-normalized (unitary) IFFT matrix. Its $(k,n)$ entries are given by
\begin{equation}
\{{\mathbf F}\}_{k,n}=\frac{1}{\sqrt{N}}\exp\Big(+\jmath \frac{2\pi k}{N}n\Big),
\end{equation}
where $\jmath=\sqrt{-1}$ is the imaginary unit, $k=-\frac{N}{2},\ldots,\frac{N}{2}-1$ is the frequency-domain index, and $n=0,\ldots,N-1$ is the time-domain index.
${\mathbf W}$ is the proposed unitary precoding matrix.
Finally, ${\mathbf B}$ is an $N \times M$ matrix that maps the real and imaginary parts of the complex-valued vector ${\mathbf x}_k$ to the active subcarriers before precoding.
${\mathbf B}$ consists of $M$ columns of the identity matrix ${\mathbf I}_N$ whose indices belong to the set of active subcarriers indices ${\mathbb A}.$

\begin{figure*}[t]
    \centering
    \includegraphics[width=\textwidth]{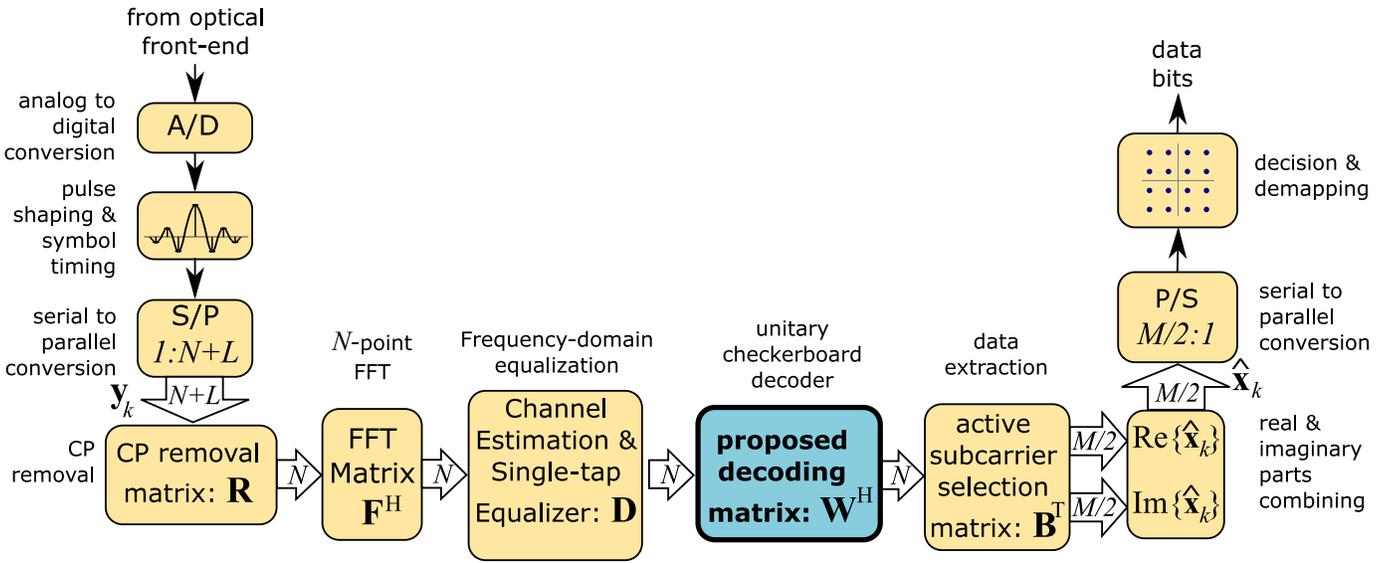}
    \caption{RX block diagram. The proposed decoder is the highlighted block.}
    \label{fig:RX}
\end{figure*}

\subsection{Receiver Architecture}

The receiver (RX) block diagram is shown in Figure~\ref{fig:RX}.
The analog waveform received from the OWC channel via the optical front-end is digitized and pulse-shaped.
Then, after OFDM symbol timing recovery, the serial sequence is down-sampled, and converted to vectors ${\mathbf y}_k$ of length $(N+L)$.
CP is discarded, and then FFT operation is required in order to perform the single-tap frequency-domain equalization by using matrix ${\mathbf D}$.
Then followed by unitary decoding (the inverse of unitary precoding), in order to obtain the estimated real and imaginary parts
of the constellation symbols  $\hat{\mathbf x}_k$ on the active subcarriers.
This is followed by serialization, symbol decision and decoding of data bits.

The overall signal processing on the horizontal part of the RX chain is described by the following simple matrix equation
\begin{equation}
\left[
\begin{array}{cc}
{\mathrm{Re}}\{\hat{\mathbf x}_k\} \\
{\mathrm{Im}}\{\hat{\mathbf x}_k\} \\
\end{array}
\right]
=
{\mathbf B}^{\textrm T}
{\mathbf W}^{\textrm H}
{\mathbf D}
{\mathbf F}^{\textrm H}
{\mathbf R}
{\mathbf y}_k,
\end{equation}
where ${\mathbf y}_k$ are $(N+L)$ vectors which contain the discretized OWC channel output, ${\mathbf R}$ is the $N \times (N+L)$ CP removal matrix,
which comprises the last $N$ rows of the $(N+L) \times (N+L)$ identity matrix ${\mathbf I}_{N+L}$.
The single-tap equalizer is represented by an $N \times N$ diagonal matrix ${\mathbf D}$
which contains the inverse of estimated channel frequency response of each active subcarrier along its diagonal, and zeros in rest.

\begin{figure}[t]
    \centering
    \includegraphics[width=0.8\columnwidth]{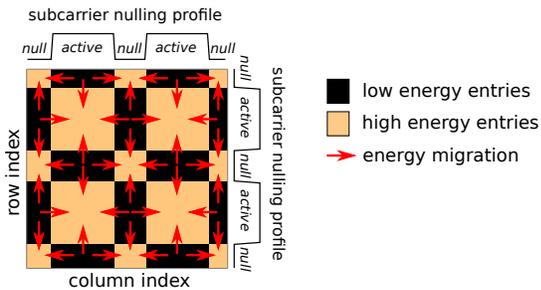}
    \caption{Energy transformation concept in the proposed UCP-OFDM precoding matrix. Energy is extracted from certain rectangular regions of the FFT matrix
      in order to preserve the locations of the null subcarriers, while conserving the overall energy. The result is a checkerboard-like matrix. }
    \label{fig:Fh}
\end{figure}

\section{Proposed Unitary Checkerboard Precoder \label{sec:precoder}}

With the three key OWC characteristics provided in Section I in mind,
the primary goal in the proposed precoder design is to achieve signal PAPR levels similar to baseband (BB) transmission while avoiding two of its deficiencies.
First, the BB modulation is extremely sensitive to DC baseline fluctuations, especially when high-order constellations are being used.
This renders it unusable in practical OWC systems, where ambient light, fluorescent lights and AC coupling in various transceiver stages cause such fluctuations, unless up/down-conversion are employed.
In addition, BB transmission does not inherently provide full control over the signal spectrum.
Hence, using the FFT matrix as a precoding matrix, which would essentially produce a single carrier signal, is unfeasible.

Our aim is to design a precoding matrix that provides full control over the null subcarriers. This allows for completely eliminating the effects of DC baseline fluctuations that BB modulation is subject to.
%Second, on-demand spectral blanking enables seamless multiple access in frequency domain.

The idea is to combine an FFT-like precoding with the IFFT-based operation in the OFDM transmission chain such that the resulting matrix is close to the identity matrix.
This way, the PAPR of the resulting OFDM signal is reduced to near the level achievable by baseband transmission (unmodulated constellation symbols),
while preserving the locations of the null subcarriers. In brief, the proposed precoded transmission is a modified BB transmission that allows for flexible subcarrier nulling (by using a subcarrier nulling filter).

The proposed precoding matrix stems from the Fast Fourier Transform (FFT) matrix, which is modified by nulling certain contiguous regions, while preserving its unitary property.
Figure~\ref{fig:Fh} illustrates the energy transformation concept by visualizing the heat map of the proposed precoding matrix.
More specifically, the FFT matrix is modified by acting on the magnitude of several entries that are grouped in rectangular regions.
The energy is redistributed among these rectangular regions of the matrix while preserving the total energy.
The resulting matrix exhibits a checkerboard-like pattern.
Imposing such a pattern on the precoding matrix is {\em essential} since it allows for the preservation of the null subcarriers.
Since this pattern is at the core of the proposed precoder, we call the proposed precoding method ``Unitary Checkerboard Precoded OFDM'', henceforth simply abbreviated as UCP-OFDM.

\begin{figure}[t]
    \centering
    \includegraphics[width=0.7\columnwidth]{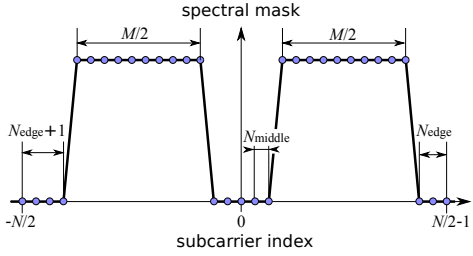}
    \caption{ Illustration of active/null subcarrier spectral mask ${\mathbf m}$ with $M=20$, $N_{\text{middle}}=2$  and $N_{\text{edge}}=3$ }
    \label{fig:mask}
\end{figure}

Let us first define the binary active/null subcarrier spectral mask ${\mathbf m}$ as an $N \times 1$ vector which allows for a flexible subcarrier nulling pattern.
Its entries are either one, if the corresponding subcarrier index belongs to the set of active subcarriers, which we denote by ${\mathbb A}$, or zero for the null subcarriers.
The $k$-th entry of vector ${\mathbf m}$ is given by
\begin{equation} \label{m}
{\mathbf m}_k=
\left\{
\begin{array}{cc}
1, & \mbox{if} \: k \in {\mathbb A} \\
0, & \mbox{otherwise}
\end{array}.
\right.
\end{equation}
In most practical OFDM transceivers, the subcarrier in the middle of the spectrum, as well as the subcarrier at the lower edge of the band are unmodulated (null).
The middle subcarrier is null to avoid the unstable baseline voltage of the OWC signal,
whereas the lower edge subcarrier is null to preserve the spectrum symmetry (give the even FFT size, usually a power of 2).
We further generalize the idea of null subcarrier control by assuming that there are $N_{\text{middle}}$ additional null subcarriers on each side of the middle subcarrier,
$N_{\text{edge}}+1$ null subcarriers at the lower edge of the band and $N_{\text{edge}}$ null subcarriers at the upper edge of the band.
Therefore, the total number of active subcarriers is $M=N-2(N_{\text{middle}}+N_{\text{edge}}+1)$.
An example of active/null subcarrier spectral mask ${\mathbf m}$ is illustrated in Figure~\ref{fig:mask}, assuming $N=32$ subcarriers, of which $M=20$ are active.
There are $N_{\text{middle}}=2$ null subcarriers on each side of the middle subcarrier, $N_{\text{edge}}+1=4$ null subcarrier at the lower edge of the spectrum,
and $N_{\text{edge}}=3$ null subcarriers at the upper edge.
Such a null subcarrier assignment is sufficiently general for most practical OFDM systems, not only OWC.
For example, in Wi-Fi (the IEEE 802.11a/g/n standard), $N=64$, $M=52$, $N_{\text{middle}}=0$ and $N_{\text{edge}}=5$.
However, this particular type of spectral mask is {\em not} a strict requirement for the proposed precoder.
As long as spectrum symmetry is preserved, null and active subcarriers can be assigned arbitrarily, for example, to enable multi-user access.

In order to be able to ensure the desired subcarrier nulling, let us define the two-dimensional spectral mask matrix:
\begin{equation} \label{M}
{\mathbf M}={\mathbf m}{\mathbf m}^{\mathrm T}
+ ({\mathbf 1}_N-{\mathbf m})({\mathbf 1}_N-{\mathbf m})^{\mathrm T},
\end{equation}
where ${\mathbf 1}_n$ is an $n \times 1$ vector of ones, and $(\cdot)^{\mathrm T}$ denotes the matrix transpose.
The structure of matrix ${\mathbf M}$ exhibits a special binary pattern which looks like a checkerboard, as shown in Figure~\ref{fig:Fh}.
%Figure~\ref{fig:mask2d},
%where dark color represents zero entries, whereas light color represents unit entries.
%
%\begin{figure}
%    \centering
%    \includegraphics[width=\columnwidth]{M_matrix}
%    \caption{Checkerboard-like heat map of the binary spectral mask matrix ${\mathbf M}$ with $M=20$, $N_{\text{middle}}=2$  and $N_{\text{edge}}=3$.}
%    \label{fig:mask2d}
%\end{figure}
From now on, we will call a {\em patterned matrix}, any matrix whose zero entries exhibit the same structure as the zeros in matrix ${\mathbf M}$ in~\eqref{M},
and has arbitrary entries in rest (in place of the unit elements in ${\mathbf M}$).
Similarly, we further call a {\em patterned vector}, any vector whose zero entries exhibit the same structure as the zeros in vector ${\mathbf m}$ in~\eqref{m},
and has arbitrary entries in rest.
Since rows and columns of any patterned matrix whose indices correspond to null subcarriers are orthogonal to any patterned vector of matching dimensions,
this pattern allows the preservation of the null subcarriers.
Therefore, is important to impose the checkerboard pattern on the proposed precoder.
The multiplication of a patterned matrix by a patterned vector produces another patterned vector. Hence locations of null subcarriers are preserved.
A patterned precoding matrix may be obtained, for example, by multiplying element-wise an arbitrary square matrix by the checkerboard-like spectral mask matrix ${\mathbf M}$.
However, this needs to be invertible, in order to ensure perfect symbol recovery.

\begin{figure}[t]
    \centering
    \includegraphics[width=\columnwidth]{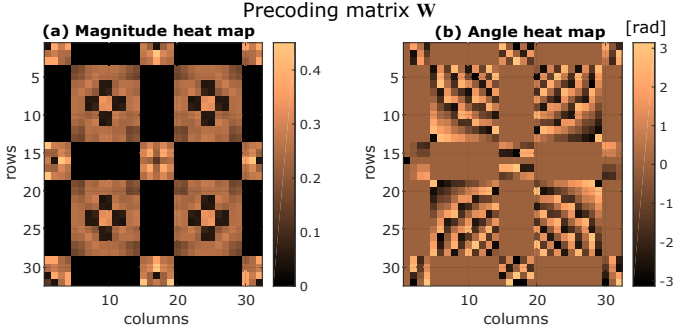}
    \caption{ Heat map of magnitude and angle of the elements of the proposed UCP-OFDM precoding matrix ${\mathbf W}$. }
    \label{fig:W}
\end{figure}

The proposed precoding matrix ${\mathbf W}$ is chosen to satisfy the following criteria:
\begin{itemize}
  \item[1.] be as close as possible in Frobenius norm to the FFT matrix, in order to produce very low PAPR (like the FFT-precoded OFDM)
  \item[2.] be a patterned matrix in order to preserve the location of null subcarriers.
  \item[3.] be unitary (like the FFT matrix), in order to avoid noise coloring, and at the same time, ensure full rank for lossless symbol recovery at the receiver (RX).
\end{itemize}
Therefore, we formulate the problem of finding the precoding matrix as a matrix optimization problem under the constraint that ${\mathbf W}$ is an $N \times N$ unitary patterned matrix, i.e.:
\begin{equation}
{\mathbf W}=\arg\min_{\mathbf Q}
\{\|{\mathbf Q}
-
{\mathbf F}^{\mathrm H} \odot {\mathbf M}\|_{\mathrm F}^2
\}
\end{equation}
subject to ${\mathbf W}^{\text{H}}{\mathbf W}={\mathbf I}_N$ and ${\mathbf W}={\mathbf W} \odot {\mathbf M},$
where $\|\cdot\|_{\mathrm{F}}$ represents the matrix Frobenius norm, $\odot$ denotes the Hadamard (element-wise) matrix product, and $(\cdot)^{\mathrm H}$ is the Hermitian transpose of a matrix.
The solution to the above problem is the orthogonal projection (under the standard Euclidean metric) of the patterned matrix ${\mathbf F}^{\mathrm H} \odot {\mathbf M}$
onto the Lie group of $N \times N$  unitary matrices ${\mathbb{U}}(N)$~\cite{Hig89,2008_Abr_PhD_thesis}.
The orthogonal projection of an arbitrary matrix ${\mathbf A}$ onto ${\mathbb{U}}(N)$ can be obtained from its singular value decomposition
${\mathbf A}={\mathbf U}{\boldsymbol\Sigma}{\mathbf V}^{\textrm H}$ as ${\text{proj}}_{{\mathbb{U}}(N)}\{{\mathbf A}\}={\mathbf U}{\mathbf V}^{\textrm H}$~\cite{2008_Abr_PhD_thesis}.
Therefore, the desired precoder matrix is obtained as:
\begin{equation} \label{W}
{\mathbf W}={\text{proj}}_{{\mathbb{U}}(N)}
\{
{\mathbf F}^{\mathrm H} \odot {\mathbf M}
\}
\end{equation}
The result of this projection is a patterned unitary precoding matrix ${\mathbf W},$ as required.
The matrix ${\mathbf W}$ in~\eqref{W} is essentially a modified FFT matrix ${\mathbf F}^{\mathrm H}$
such that the locations of the null subcarriers defined by the vector ${\mathbf m}$ are preserved after multiplying by it.
The modification is minimal as measured by Frobenius norm, and therefore, the proposed scheme may be seen as a quasi-baseband modulation.
The heat map of magnitude and angle of its elements are illustrated in Figure~\ref{fig:W}.
The zero entries obey the imposed checkerboard-like pattern, as shown by the magnitude heat map in Figure~\ref{fig:W}(a).
In addition, similarly to the FFT matrix, ${\mathbf W}$ exhibits Hermitian-symmetry w.r.t the index of the middle subcarrier, as seen in Figure~\ref{fig:W}(b).
Consequently, after applying the IFFT operation, this property ensures that the resulting combined precoding-modulation matrix is {\em real-valued}.
Then, since our composite precoding-modulation matrix is real-valued, the input signal must be real-valued as well, in order to obtain a real-valued baseband OWC signal.
This is the reason why the complex constellation needs to be split into real and imaginary parts prior to precoding.

%%%%%%%%%%%%%%%%%%%%%%%%%%%%%%%%%%%%%%%%%%%%%%%%%%%%%%%%%%%%%%%%%%%%%%%%%%%%%%%%%%%%%%%%%%%%%%%%%%%%%%%%%%%%%%%%%%%%%%%%%%%%%%%%%%%%%%%%%%%%%%

\section{Implementation}
\label{implementation}
In this section we address the computational complexity and storage required in order to implement the proposed UCP-OFDM transceiver. Numerical issues are also briefly addressed.

\subsection{Implementation of Precoding and Modulation}

First, we may notice in Figure~\ref{fig:TX} that on the TX side, both IFFT and multiplication by full-rank ${\mathbf W}$ precoding matrix are required.
Therefore, the direct implementation of the proposed precoder would exhibit complexity of order $N^2$, i.e. squared in the number of subcarrier,
which for large $N$, would be much higher compared to the complexity $N \log_2 N$ of the radix-2 FFT/IFFT operations in the conventional OFDM transceiver.
Apart from that, an additional IFFT operation would still be required. For this reason, the precoding matrix ${\mathbf W}$ should not be implemented separately.

In order to circumvent this complexity issue, we exploit the fact that ${\mathbf W}$ is close (in Frobenius norm) to the IFFT matrix. Therefore, the composite precoding-modulation matrix
which consists of the multiplication of IFFT and the proposed precoding matrix ${\mathbf W}$, defined as
\begin{equation}
{\mathbf P} = {\mathbf F} {\mathbf W}
\end{equation}
is close to the $N \times N$ identity matrix ${\mathbf I}_N$. The composite precoding-modulation matrix ${\mathbf P}$ has the following three {\em key} properties:
\begin{itemize}
  \item[(P1)] It is unitary, i.e., ${\mathbf P}^{\textrm H}{\mathbf P}={\mathbf I}_N$, being a product of two unitary matrices.
  \item[(P2)] Most of its eigenvalues are equal to one, due to its nearness to identity matrix ${\mathbf I}_N$.
  \item[(P3)] It is real-valued (hence orthogonal), since IFFT is applied to ${\mathbf W}$, whose columns exhibit Hermitian symmetry.
\end{itemize}
We exploit the three properties above in order to achieve a computationally efficient implementation solution based on Singular Value Decomposition (SVD).
Let us define the matrix
\begin{equation}
{\mathbf E}={\mathbf P}-{\mathbf I}_N
\end{equation}
Due to property (P2) above, ${\mathbf E}$ exhibits low-rank. Its compact (rank-$r$) SVD is given by
\begin{equation} \label{E}
{\mathbf E}={\mathbf U}_{\text{r}}{\boldsymbol\Sigma}_{\text{r}}{\mathbf V}_{\text{r}}^{\textrm H}
\end{equation}
Therefore, the efficient implementation of the composite precoding-modulation matrix ${\mathbf P}$ is as follows:
\begin{equation} \label{IN+E}
{\mathbf P}={\mathbf I}_N + \underbrace{{\mathbf U}_{\text{r}}{\boldsymbol\Sigma}_{\text{r}}{\mathbf V}_{\text{r}}^{\textrm H}}_{\mathbf E}
\end{equation}
We may notice in~\eqref{IN+E} that the composite precoding-modulation matrix ${\mathbf P}$ is expressed as the $N \times N$ identity matrix ${\mathbf I}_N$ plus the SVD of low-rank matrix ${\mathbf E}$.
In other words, the output signal consists of the original baseband signal plus a {\em subcarrier nulling signal} that ensures the desired null subcarrier pattern defined by vector $\mathbf m$.
The subcarrier nulling signal is obtained by multiplying the original baseband signal by matrix ${\mathbf E}$ (written in SVD form), which we call {\em subcarrier nulling matrix}.

\begin{figure}[t]
    \centering
    \includegraphics[width=0.6\columnwidth]{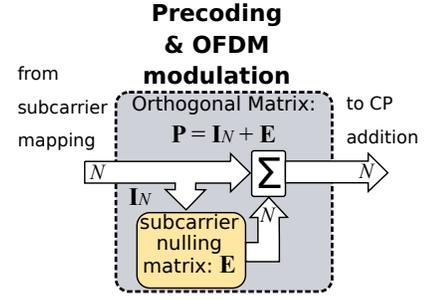}
    \caption{Efficient implementation of the proposed UCP as a quasi-baseband transmission via SVD.}
    \label{fig:TX_implementation}
\end{figure}

The equivalent block diagram of the combined precoding-modulation matrix ${\mathbf P} = {\mathbf F} {\mathbf W}$, which contains both the proposed UCP and the IFFT matrix
is given in Figure~\ref{fig:TX_implementation}. We may notice the quasi-BB nature of our proposed transmission scheme, the main (upper) branch corresponds to the original BB signal,
whereas the secondary (lower) branch corresponds to the subcarrier nulling signal. In this implementation, no FFT/IFFT operations are required on the TX side (IFFT operation is moved on the RX side).

The subcarrier nulling matrix ${\mathbf E}$ exhibits low rank $r=2Z$, Thus, precoding only requires $2NZ$ scalar real-valued multiplications/additions.
In general, $2Z<\log_2 N$, thus the proposed precoding-modulation operation is computationally simpler than radix-2 FFT/IFFT operations used in conventional OFDM transceivers,
which amount for $N \log_2 N$.
In most practical OFDM-based OWC transceivers, the middle subcarrier, as well as the subcarrier at the lower edge of the band are unmodulated (null).
Hence, $Z=2$, and the number of operations required by the precoding-modulation would only be $4N$, i.e., linear complexity in number of subcarriers.
In terms of required memory, the $N \times 2Z$ real-valued matrix multiplication result ${\mathbf U}_{\text{r}}{\boldsymbol\Sigma}_{\text{r}}$
and the $N \times 2Z$ real-valued matrix ${\mathbf V}_{\text{r}}$ need to be stored. If $Z=2$, this only amounts to $8N$ real-valued numbers. 

\subsection{Demodulation and Decoding Implementation}

It is important to mention that ${\mathbf P}$ is a real-valued orthogonal matrix, i.e., ${\mathbf P}^{\textrm T}{\mathbf P}={\mathbf I}_N$.
Therefore, a real-valued input signal produces a real-valued output signal. Decoding is trivial, the symbol recovery is achieved via multiplication by its transpose ${\mathbf P}^{\textrm T}$.
The real-valued orthogonal decoder (which jointly performs the OFDM demodulation and undoes the precoding operation) is very similar to the encoder in Figure~\ref{fig:TX_implementation},
with the only exception that ${\mathbf E}^{\textrm T}$ needs to be used instead of ${\mathbf E}$.
Again, in order to achieve ultra-low implementation complexity, the demodulation and decoding is achieved relying on the SVD representation of ${\mathbf E}$.
\begin{equation}
{\mathbf P}^{\textrm T}={\mathbf I}_N + \underbrace{{\mathbf V}_{\text{r}}{\boldsymbol\Sigma}_{\text{r}}{\mathbf U}_{\text{r}}^{\textrm T}}_{{\mathbf E}^{\textrm T}}
\end{equation}

On the receiver side, both FFT and IFFT are required in order to perform the trivial single-tap frequency-domain equalization.
This is a way to avoid the multiplication by the decoding matrix ${\mathbf W}^{\mathrm H}$ alone, whose complexity would be of order $N^2$.
Instead of multiplying by ${\mathbf W}^{\mathrm H}$, the equivalent multiplication by ${\mathbf P}^{\textrm{T}} {\mathbf F}$ is used, whose complexity is much lower, i.e., $N \log_2 N + 2NZ$
(demodulation-decoding exhibits lower complexity than FFT, i.e., $2NZ$).
Single-tap channel estimation and equalization require $M$ operations each (as they are only applied to active subcarriers).
Thus, the benefit of simple channel estimation and equalization fully justifies moving the IFFT from the TX side to the RX side.
This difference becomes less significant if radix-4 FFT is used.
In conclusion, the overall computational complexity of the proposed receiver that includes single-tap channel estimation, frequency-domain channel equalization,
OFDM demodulation and symbol decoding is of order of $2N \log_2 N + 2NZ + 2M$ scalar operations (assuming radix-2 FFT), very similar to the conventional OFDM.
The amount of memory required at RX is the same as for TX.

\section{Evaluation}
\label{evaluation}
In this section, we assess the performance of the proposed UCP-OFDM transmission scheme w.r.t state-of-the-art methods.
In Section~\ref{sec:PAPR} we analyze the PAPR levels of different modulation schemes.
Resilience against low-frequency disturbances is shown in Section~\ref{sec:DC_wander}.
%The subcarrier nulling capability of the proposed scheme is demonstrated in Section~\ref{sec:subcarrier_nulling}.
In Section~\ref{sec:BER}, we compare the bit error rate performance of the proposed UCP-OFDM method to other modulation schemes.

\subsection{PAPR Analysis \label{sec:PAPR}}

\begin{figure}[t]
    \centering
    \includegraphics[width=0.8\columnwidth]{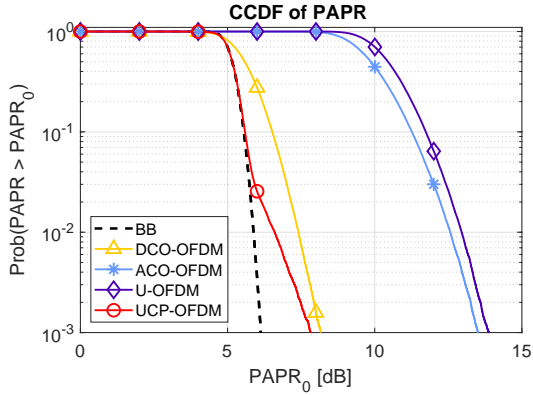}
    \caption{The proposed OFDM precoding scheme outperforms state-of-the-art schemes in terms of PAPR,
      and is just 1--2 dB above the level of baseband modulation in 99.9\% of time.
      \label{fig:PAPR_CCDFs}}
\end{figure}

In this section, we assume a multicarrier OWC transmission with $N=256$ subcarriers of which $M=254$ are active.
The null subcarriers are the middle (DC) subcarrier, as well as one subcarrier at the lower edge of the spectrum.
We analyze the PAPR of different transmission schemes after pulse shaping and oversampling,
which are performed by using a poly-phase root-raised cosine filter with oversampling ratio of 8, roll-off factor of 0.25, and group delay of 8 samples at the lower rate. We assume a 16-QAM constellation. DC-biasing is also applied after pulse-shaping in the case of bipolar schemes (DCO-OFDM and UCP-OFDM), prior to PAPR computation.
The PAPR complementary cumulative distribution function (CCDF) of the pulse-shaped time-domain signals is shown in Figure~\ref{fig:PAPR_CCDFs}.
The following relevant schemes are considered in the comparison:
BB modulation, which can be considered a lower bound in terms of PAPR, but is unsuitable for OWC systems unless up/down-conversion are employed,
the classical DCO-OFDM, as well as two unipolar schemes, ACO-OFDM and U-OFDM.

We may notice in Figure~\ref{fig:PAPR_CCDFs} that the proposed UCP-OFDM achieves PAPR levels which are just about 1-2 dB higher than the levels corresponding to a baseband signal,
assuming the same constellation type.
However, unlike BB, the proposed scheme offers flexible subcarrier nulling to completely mitigate the low frequency disturbances.
Due to the quasi-BB nature of UCP-OFDM signal, when lower order constellations are being used, PAPR levels are even lower,
unlike the other schemes, for which the constellation type does not impact much the PAPR levels, due to the large number of subcarriers.
The PAPR gap between the proposed UCP-OFDM and DCO-OFDM is not significant.
However, the bit error rate performance of the proposed scheme is superior to DCO-OFDM, the SNR gain in multipath channels is about 5 dB,
and higher compared to ACO-OFDM and U-OFDM subject to same throughput, as it will be shown later in Section~\ref{sec:BER}.

%\subsection{Subcarrier Nulling \label{sec:subcarrier_nulling}}
%The proposed scheme allows for on-demand subcarrier nulling, unlike BB modulation, thus avoiding the low %frequency disturbances the OWC signal is subject to.
%In order to demonstrate this feature, we assume $N=256$ subcarriers, of which $Z=6$ are null subcarriers.
%There are 5 null subcarriers in the middle of the spectrum (2 on each side of the DC subcarrier, which is also null), and one at the edge of the spectrum.
%The averaged amplitude spectrum of the oversampled precoded OFDM signal is shown in Figure~\ref{fig:spectrum}.
%Deep subcarrier nulling (40 to 50 dB) is achieved, as seen in Figure~\ref{fig:spectrum}.
%Different subcarrier nulling profiles may be used, for example, in order to enable multi-user frequency domain multiplexing.

%\begin{figure}[t]
%    \centering
%    \includegraphics[width=\columnwidth]{spectrum}
%    \caption{ Amplitude spectrum of the proposed precoded OFDM signal. There are five null subcarriers in the middle of the spectrum and one near the edge of the spectrum.}
%    \label{fig:spectrum}
%\end{figure}

\subsection{Resilience against low-frequency disturbances: UCP-OFDM vs. BB modulation \label{sec:DC_wander}}

In general, low frequency disturbances manifest as slow fluctuations in baseline voltage (baseline wander).
They may be caused by variations of ambient light in the vicinity of the receiver (especially in mobile scenarios),
flickering effect from fluorescent lights~\cite{2016_IslHaa,2014_TsoVidHaa}, as well as
a DC-imbalanced waveforms undergoing AC-coupling in the OWC front-end~\cite{2013_AzhBri,2016_GroJunLanHaaWol} (loss of lower frequency components).
In general, baseline wander is not an issue for DCO-OFDM, as shown experimentally in~\cite{2016_GroJunLanHaaWol}.
However, unlike DCO-OFDM, baseband signals~\cite{2014Wan_etal,2016_LinHuaJi}, are unable to avoid low-frequency spectrum, unless up-conversion is employed, as in~\cite{2014Wan_etal}.
Furthermore, practical experiments in~\cite{2016_GroJunLanHaaWol} demonstrate that baseline wander was a problem not only for baseband, but also for some unipolar OFDM waveforms.
Both ACO-OFDM and U-OFDM suffer from baseline wander issue due to the zero-level clipping in presence of bandwidth limitation, which lead to severe bit-rate degradation.
Consequently, they require a additional DC biasing. In summary, the low-frequency disturbances should not be ignored in any practical OWC design, including some OFDM-based designs.

\begin{figure}[t]
    \centering
    \includegraphics[width=0.8\columnwidth]{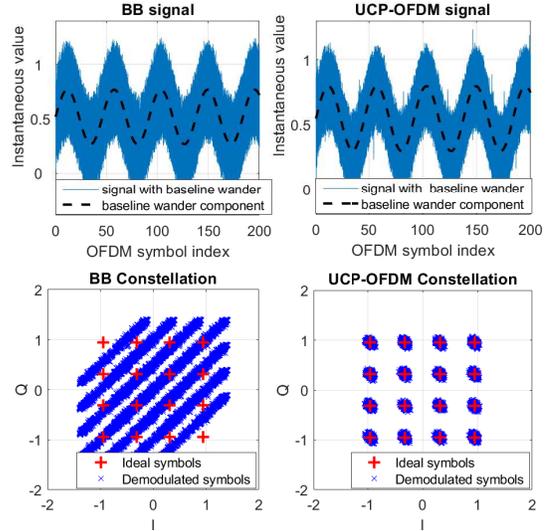}
    \caption{The impact of signal baseline wander on the BB modulation (left hand side) vs. the proposed UCP-OFDM (right hand side).
      The time-domain signals affected by baseline wander are shown in the upper subplots.
      The resilience to baseline fluctuations is clearly visible in the two lower plots showing the corresponding demodulated constellations.}
    \label{fig:DC-wander}
\end{figure}

In this section, we demonstrate the robustness of the proposed UCP-OFDM scheme against signal baseline fluctuations.
We vary the signal baseline according to a sinusoidal function whose root-mean-square (RMS) value equals the time-domain waveform standard deviation,
and whose period spans about 45 OFDM symbols. Such fluctuations may be caused, for example, by fluorescent lights (flickering), or parasite harmonic signals captured from the environment
(low-frequency electro-magnetic interference). 
We compare the proposed quasi-BB UCP-OFDM scheme with conventional BB scheme by showing the demodulated constellations
corresponding to a time interval of 200 OFDM symbols, assuming AWGN channel and noise power of $-40$ dB.
Figure~\ref{fig:DC-wander} shows the time-domain waveforms (top) and the corresponding demodulated constellations (bottom), for BB (left) and UCP-OFDM (right), respectively.
The DC baseline fluctuations produce high interference among the demodulated BB symbols, whereas the UCP-OFDM constellation points remain clearly separated.
The resilience of the proposed UCP-OFDM against low frequency disturbances is enabled by nulling the middle subcarriers (in this case the DC subcarrier only),
which is not achievable in the conventional BB modulation.

\subsection{Bit Error Rate Analysis \label{sec:BER}}

In this section, we evaluate the bit error rate (BER) performance of the proposed UCP-OFDM transmission scheme vs state-of-the-art schemes such as DCO-OFDM, ACO-OFDM, and U-OFDM.
We have implemented eACO-OFDM scheme as well, but unfortunately, we were only able to get it to work in AWGN channels,
the BER performance degraded fast in multipath channels.
In this evaluation, we do not include coded schemes such as LACO-ODFM~\cite{2020_PurYesSafHaa} and SC-GTIM~\cite{2021_ZhaZunPetHaa}. 
We assume that all multicarrier transmission schemes use $N=256$ subcarriers that span a bandwidth of 625 MHz, and the CP length is $L=16$,
which can accommodate a maximum delay spread value of about $30$~ns.
The transmitted packets consist of $P=10$ OFDM symbols preceded by a preamble symbol that is used for channel estimation.
The frequency-domain training sequence is a Zadoff-Chu sequence with Hermitian symmetry, in order to ensure a real-valued time-domain output sequence.
The channel corresponding to every packet is estimated from the corresponding preamble, and single-tap frequency-domain channel equalization is performed for all methods by using a zero-forcing equalizer.

\begin{figure}[t]
    \centering
    \includegraphics[width=\columnwidth]{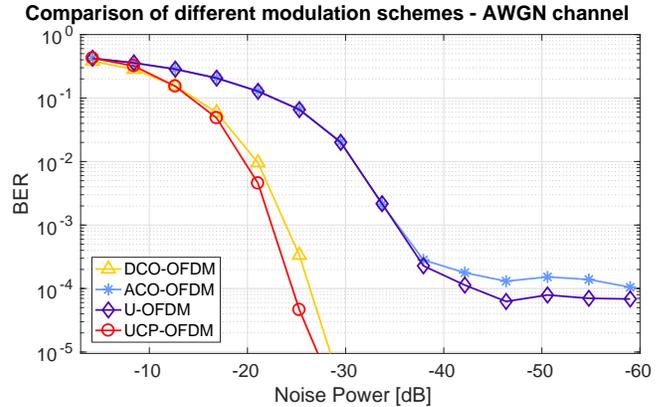}
    \caption{Bit error rate comparison in AWGN channels.}
    \label{fig:BER_AWGN}
\end{figure}

\begin{figure}[t]
    \centering
    \includegraphics[width=\columnwidth]{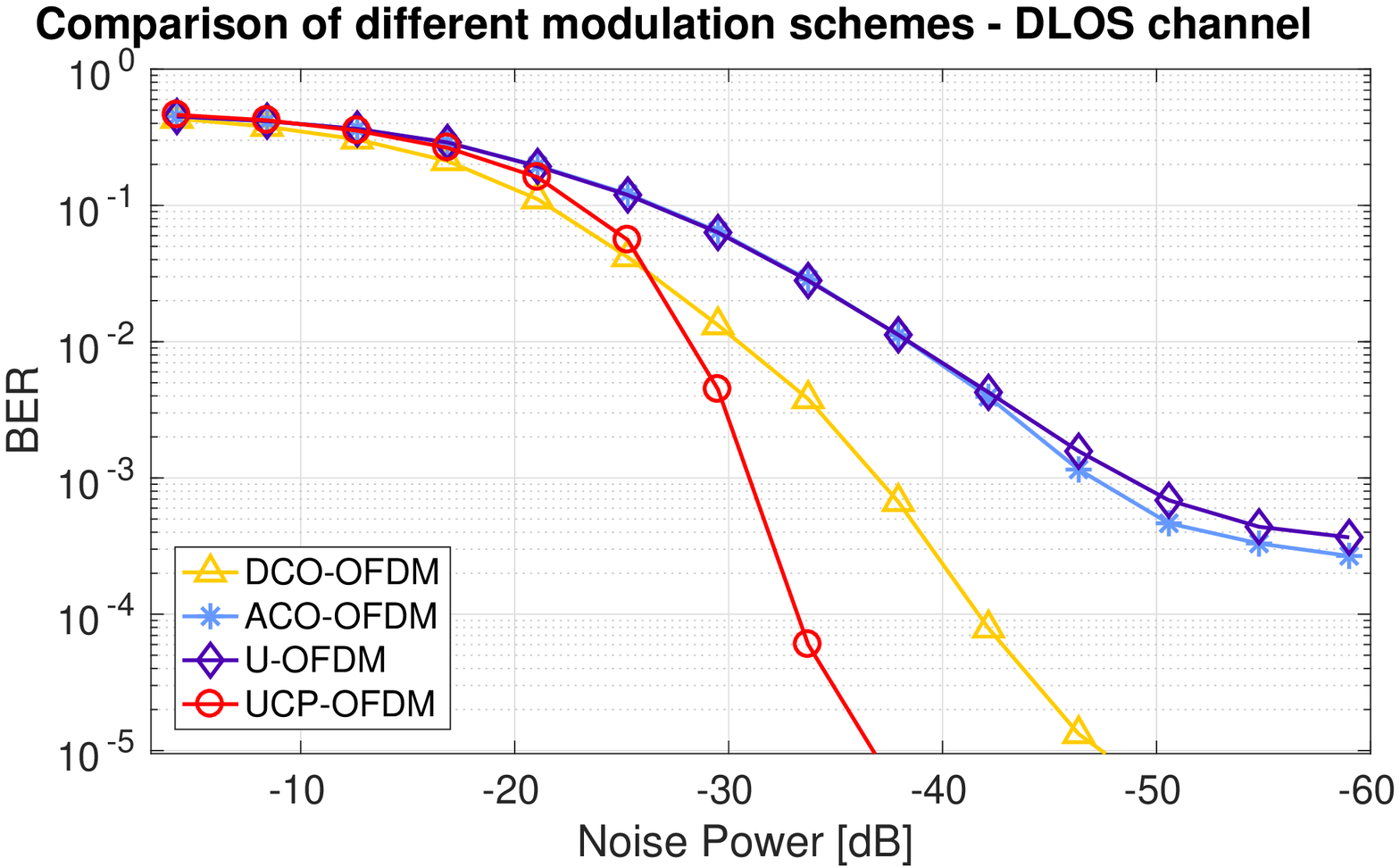}
    \caption{Bit error rate comparison in DLOS multipath channels.}
    \label{fig:BER_DLOS}
\end{figure}

\begin{figure}[t]
    \centering
    \includegraphics[width=\columnwidth]{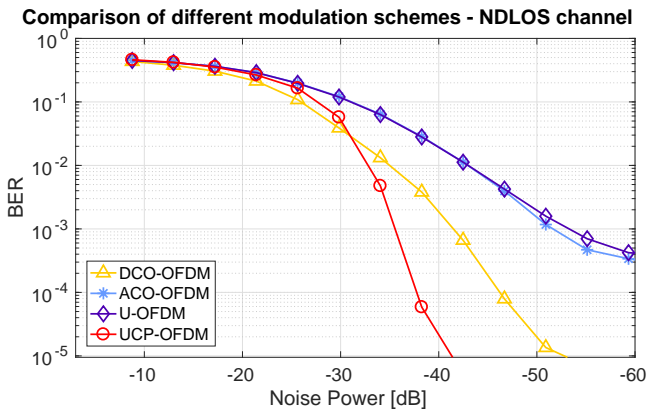}
    \caption{Bit error rate comparison in NDLOS multipath channels.}
    \label{fig:BER_NDLOS}
\end{figure}

In order to have a fair comparison, we fix the dynamic range of the TX signal to be the same for all methods, according to the TX front-end limitations
(here, without loss of generality, the lower and upper limit are 0 and 1, respectively).
The receiver noise figure is set to be the same for all schemes.
We separately optimize the transmit power of every scheme, in order to minimize the BER in the moderate SNR region (around 20 dB).
  Hence each scheme will experience different clipping probabilities, such that the signal amplitude is optimally scaled within the available dynamic range.
  The optimal clipping probabilities for the unipolar schemes ACO-OFDM and U-OFDM are quite mild, $0.69 \times 10^{-3}$ and $0.97 \times 10^{-3}$, respectively.
  DCO-OFDM and the proposed UCP-OFDM tolerate more severe clipping, the optimal clipping probabilities are $4.4 \times 10^{-2}$ and $2.2 \times 10^{-2}$,
  respectively, which ultimately results into higher average transmitted power.

Unipolar signals (ACO-OFDM, and U-OFDM) are clipped only on the positive limit, whereas bipolar signals (DCO-OFDM and UCP-OFDM) are being clipped on both sides.
In addition, for the full-rate schemes (DCO-OFDM and UCP-OFDM), we assume 16-QAM input constellation.
In order to maintain the same throughput, for the half-rate schemes (ACO-OFDM, and U-OFDM),
the number of bits per constellation symbol is doubled (e.g. 256-QAM is being used), as also suggested in~\cite{2015_IslTsoHaa}.
We compared the BER performance of the schemes by considering three types of OWC channels:
additive white Gaussian (AWGN), where only noise is present, and channel is a unit impulse, directed line-of-sight (DLOS) multipath channel, where the LOS component dominates over multipath,
as well as non-directed line-of-sight (NDLOS) multipath channel, where user is slightly beyond the cell edge, and multipath dominates over the LOS component.
The multipath channel model is derived from~\cite{2012_GhaPopRaj} and~\cite{2012_BurMinGha_etal_1}, which provided the way of calculating the DC channel gains of the direct path and reflected paths. An office environment channel is considered, and the link-related parameters are summarized in Table~\ref{sim_params}.
A number of 1000 independent runs are performed, each having a different channel realization, which for fairness, is identical to all methods.
In each run, a number of 5 packets is transmitted (50 OFDM symbols). The BER is averaged over all runs, for each type of OWC channel.

\begin{table}[t]
\centering
\small
\begin{tabular}{|c|p{32mm}|}
\hline
{Total Transmitted Power}:  & 20 W \\ \hline
{Angle of irradiance in half } & $30^\circ$ \\ \hline
{FOV (field of view) of detector in half} & $60^\circ$ \\ \hline
{Detector area} & $7.8\times 10^{-7} $m$^2$ \\ \hline
{PD Concentrator refractive index} & 1.46 \\ \hline
{Optical filter gain} & 1 dB \\ \hline
{Optical concentrator gain} & 1 dB \\ \hline
{Reflectivity of walls} & 0.7 \\ \hline
{Room size} & 5 m $\times$ 5m $\times$ 3 m \\ \hline
{Height of transmitter} & 1.8 m \\ \hline
{Coordinate of transmitter} & (0,0) \\ \hline
{Coordinate of receiver} & (0,0) for DLOS;

(1.5,1.5) for NDLOS. \\ \hline
\end{tabular}
\caption{The parameters of simulation in Figures~\ref{fig:BER_DLOS} and~\ref{fig:BER_NDLOS}. \label{sim_params}}
\end{table}

Figure~\ref{fig:BER_AWGN} shows the BER performance of the different schemes in AWGN channel vs. receiver noise power in dB, as $P_N=20 \log_{10} \sigma$, where $\sigma$
is the noise standard deviation.
We may notice that the proposed UCP-OFDM scheme outperforms all the other schemes. There is a gain of less than 1 dB w.r.t. DCO-OFDM, and 5 dB or more w.r.t.
the unipolar schemes ACO-OFDM and U-OFDM.
In addition, the unipolar schemes experience quickly an error floor, which is caused by the clipping, combined with the higher order constellation used compared to the other two schemes.

In Figure~\ref{fig:BER_DLOS}, we compare the methods in DLOS multipath channels. We may notice a performance degradation for all methods.
However, the gain of the proposed UCP-OFDM scheme vs. DCO-OFDM becomes significant in the high SNR region (about 5 dB).
The performance of the unipolar schemes degrades further in the presence of multipath.

Figure~\ref{fig:BER_NDLOS} shows the BER performance of the four considered modulation schemes in NDLOS multipath channels.
The performance gap between the proposed UCP-OFDM and DCO-OFDM is around 5 dB, and much higher w.r.t the unipolar schemes ACO-OFDM and U-OFDM, subject to equal throughput.

In conclusion, the superior properties of the proposed schemes are manifold: ultra-low PAPR, full throughput, total control over the null subcarriers,
better BER performance in multipath channels,
as well as low computational complexity.
To date, we are not aware of any other methods to perform well w.r.t. all the above criteria.

%%%%%%%%%%%%%%%%%%%%%%%%%%%%%%%%%%%%%%%%%%%%%%%%%%%%%%%%%%%%%%%%%%%%%%%%%%%%%%%%%%%%%%%%%%%%%%%%%%%%%%%%%%%%%%%%%%%%%%%%%%%%%%%%%%%%%%%%%%%%%%%%%%%%%%%%%%%%%%%%
%{Conclusion}
\section{Conclusions \label{sec:conclusion}}

The main idea of this contribution consists in combining the IFFT operation in the traditional OFDM transmitter chain with a novel precoder derived from the IFFT matrix to obtain a quasi-baseband modulation. Thanks to this measure, ultra-low PAPR is achieved at the level comparable to the baseband modulation. From a practical perspective is important that this effect was achieved without sacrificing any bandwidth -- the proposed scheme offers full throughput without employing any undesirable redundancy. By exploiting several properties of the proposed precoder, we also ensure that subcarriers can be blanked on demand without sacrificing the overall performance. This feature is considered as a key precursor for DC subcarrier nulling as well as multi-user frequency domain multiplexing. From a practical perspective, the proposed precoder can be implemented by using basic linear matrix multiplication. All of the above features are considered key enablers for ultra-high-speed optical light communications.

\section*{Acknowledgments}
The authors thank Dong Fang for his valuable technical input.
%%%%%%%%%%%%%%%%%%%%%%%%%%%%%%%%%%%%%%%%%%%%%%%%%%%%%%%%%%%%%%%%%%%%%%%%%%%%%%%%%%%%%%%%%%%%%%%%%%%%%%%%%%%%%%%%%%%%%%%%%%%%%%%%%%%%%%%%%%%%%%%%%%%%%%%%%%%%%%%%

%\bibliographystyle{IEEEtran}

%%%%%%%%%%%%%%%%%%%%%%%%%%%%%%%%%%%%%%%%%%%%%%%%%%%%%%%%%%%%%%%%%%%%%%%%%%%%%%%%%%%%%%%%%%%%%%%%%%%%%%%%%%%%%%%%%%%%%%%%%%%%%%%%%%%%%%%%%%%%%%%%%%%%%%%%%%%%%%%%

\end{document}